\documentclass[12pt]{article}
\usepackage{graphicx,amsmath,amssymb,graphicx}

\newcommand\pubnumber{NuPhys2016-Zhou}
\newcommand\pubdate{\today}

\def\napoli{Institute for Particle Physics Phenomenology, Department of Physics, \\ 
Durham University, Durham DH1 3LE, United Kingdom}
\def\support{\footnote{Work supported by ERC Grant “NuMass” (FP7-IDEAS-ERC ERC-CG 617143)}}

\def\Title#1{\begin{center} {\Large #1 } \end{center}}
\def\Author#1{\begin{center}{ \sc #1} \end{center}}
\def\Address#1{\begin{center}{ \it #1} \end{center}}

\newcommand\pubblock{\rightline{\begin{tabular}{l} \pubnumber\\
         \pubdate  \end{tabular}}}
\newenvironment{Abstract}{\begin{quotation}  }{\end{quotation}}
\newenvironment{Presented}{\begin{quotation} \begin{center} 
             PRESENTED AT\end{center}\bigskip 
      \begin{center}\begin{large}}{\end{large}\end{center} \end{quotation}}
\def\Acknowledgements{\bigskip  \bigskip \begin{center} \begin{large}
             \bf ACKNOWLEDGEMENTS \end{large}\end{center}}

\def\beq{\begin{equation}}
\def\eeq#1{\label{#1}\end{equation}}
\def\eeqn{\end{equation}}

\def\beqa{\begin{eqnarray}}
\def\eeqa#1{\label{#1}\end{eqnarray}}
\def\eeqan{\end{eqnarray}}

\let\bar=\overbar

\def\Dslash{\not{\hbox{\kern-4pt $D$}}}
\def\dslash{\not{\hbox{\kern-2pt $\del$}}}

\def\msb{{\bar{\ssstyle M \kern -1pt S}}}

\begin{document}
\begin{titlepage}
\pubblock

\vfill
\Title{Flavour symmetric connections with CLFV}
\vfill
\Author{ Ye-Ling Zhou\support}
\Address{\napoli}
\vfill
\begin{Abstract}
Flavons are crucial for understanding lepton mixing in models with non-Abelian discrete symmetries. They also result in charged lepton flavour violation (CLFV) via the couplings with leptons. I emphasise that the flavon-triggered CLFV succeeds strong connections with lepton flavour mixing. Relations between branching ratios of CLFV decays and mixing angles are discussed, and CLFV sum rules are obtained. Flavons with masses around hundreds of GeV are consistent by current CLFV measurements. 

\end{Abstract}
\vfill
\begin{Presented}

NuPhys2016, Prospects in Neutrino Physics

Barbican Centre, London, UK,  December 12--14, 2016

\end{Presented}
\vfill
\end{titlepage}
\def\thefootnote{\fnsymbol{footnote}}
\setcounter{footnote}{0}

\section{Introduction}

Non-Abelian discrete flavour symmetries are directly motivated by large mixing angles measured by neutrino oscillation experiments \cite{PDG}. 
Leptonic flavour mixing is explained as the result of the group structure and irreducible representations of the symmetry. One milestone is the realisation of tri-bimaximal (TBM) mixing \cite{TBM} in $A_4$ \cite{Altarelli:2005yp,Altarelli:2005yx}. The TBM mixing predicts $\theta_{12}=35.3^\circ$ and $\theta_{23} = 45^\circ$, still consistent with current oscillation data in $3\sigma$ ranges. However, after observations of a relatively sizable $\theta_{13}$ \cite{reactor}, specific modifications to TBM should be considered. 

Flavons play a key role in flavour model constructions \cite{King:2014nza}. They gain vacuum expectation values (VEVs) with special directions and lead to the break of the flavour symmetry. Different residual symmetries may be roughly preserved in different flavon VEVs, and flavour mixing is achieved from their misalignment. The small breaking of the residual symmetries result in deviations of the mixing \cite{Pascoli:2016eld}. 

Couplings between flavons and leptons not only explain flavour mixing, but also contribute to charged lepton flavour violation (CLFV) \cite{Feruglio:2008ht,triality, Toorop:2010ex, Altarelli:2012bn, Kadosh:2013nra, Heeck:2014qea, Varzielas:2015joa, Muramatsu:2016bda, Kobayashi:2015gwa}. Strongly constrained by flavour symmetries and experimental data, flavon-triggered CLFV shows special properties identified with other new physics contributions \cite{Pascoli:2016wlt}. 
In what follows, I will discuss this phenomenology and its essential connection with flavour symmetries. For definiteness, the flavour symmetry is chosen to be $A_4$.

\section{Basic structures of $A_4$ flavour models}

Without lose of generality, I show how TBM is realized in most $A_4$ flavour models. Assuming $A_4$ is a flavour symmetry conserved at some high scale, it is broken when the energy scale decreases, but some residual symmetries are preserved, $Z_3 \subset A_4$ in the charged lepton sector, and $Z_2 \subset A_4$ in the neutrino sector\footnote{Another $Z_2'$ symmetry, which is not a subset of $A_4$, is usually preserved accidentally after $A_4$ breaking.}. The lepton mass matrices are constrained by the residual symmetries. The tri-bimaximal mixing is a result of the mismatch between the two residual symmetries. These residual symmetries are not precisely preserved. The small breaking of the residual symmetries leads to corrections to the flavour mixing, and gives rise to the non-zero $\theta_{13}$ and the Dirac CP-violating phase $\delta$. 

New scalars called flavons are necessary to be introduced in flavour models. They gain VEVs with special directions, driving non-trivial lepton mass structures and further realizing flavour mixing. In the simplest case, we need at least two $A_4$-pseudo-triplet flavons, $\varphi$ and $\chi$, one for charged leptons and the other for neutrinos. Combining flavons with leptons and the Higgs, we arrive at effective $A_4$-invariant operators as follows, 
\begin{eqnarray}
-\mathcal{L}_l &=& \frac{y_e}{\Lambda} (\overline{\ell_L} \varphi)_\mathbf{1} e_R H + \frac{y_\mu}{\Lambda} (\overline{\ell_L} \varphi)_{\mathbf{1}''} \mu_R H + \frac{y_\tau}{\Lambda} (\overline{\ell_L} \varphi)_{\mathbf{1}'} \tau_R H + \text{h.c.} + \cdots \,, \nonumber\\
-\mathcal{L}_\nu &=& \frac{y_1}{2\Lambda^2} \big((\overline{\ell_L} \ell_L^c)_{\mathbf{3}_S} \chi \big)_\mathbf{1} \tilde{H} \tilde{H} + \frac{y_2}{2\Lambda^2} (\overline{\ell_L} \ell_L^c)_\mathbf{1} \eta \tilde{H} \tilde{H} + \text{h.c.} + \cdots \,,
\label{eq:Yukawa_coupling}
\end{eqnarray}
where $\eta$ is an $A_4$-invariant scalar and the dots stand for subleading higher dimensional operators.  
Once the flavons take VEVs with the following directions
\begin{eqnarray}
\langle \varphi \rangle \propto (1,0,0)^T \,,\qquad
\langle \chi \rangle \propto (1,1,1)^T \,,
\label{eq:VEVs1}
\end{eqnarray}
in the Altarelli-Feruglio basis \cite{Altarelli:2005yx}\footnote{The directions of flavon VEVs are not unique, but basis-dependent. A basis transformation $\rho(g)\to U \rho(g) U^{-1}$ for the triplet representation $\rho(g)$ for $g\in A_4$ will change the VEVs $\langle \varphi \rangle$ and $\langle \chi \rangle$ to $U \langle \varphi \rangle$ and $U \langle \chi \rangle$, respectively. Flavour mixing, which are identified as the misalignment between different VEVs, will not be changed under this basis transformation. The follow-up discussion will be fixed in the Altarelli-Feruglio basis. In this basis, the simplest triplet is a pseudo-real one, with $\varphi_1^*=\varphi_1$, $\varphi_2^*=\varphi_3$, and $\chi_1^*=\chi_1$, $\chi_2^*=\chi_3$ being required.}, 
the charged leptons and neutrinos gain special Yukawa structures, or equivalently, mass structures as
\begin{eqnarray}
Y_l\propto M_l \propto \left(
\begin{array}{ccc}
 y_e & 0 & 0 \\
 0 & y_\mu & 0 \\
 0 & 0 &  y_\tau \\
\end{array} \right) \,,\quad
Y_\nu \propto M_\nu \propto \left(
\begin{array}{ccc}
 a+2b & -b & -b \\
 -b & 2b & a-b \\
 -b & a-b & 2b \\
\end{array} \right)\,.
\end{eqnarray}
And eventually, from them, the TBM mixing is obtained.

To be consistent with data, the residual symmetries should be broken, and corrections to the mixing must be included. Sources for the breaking include 
\begin{itemize}
\item
higher dimensional operators involving different flavons in the Yukawa couplings. For example, VEVs of the 3-dimensional products of $\chi$ and $\varphi$ take this direction $(2,-1,-1)^T$ or $(0,1,-1)^T$. They break $Z_3$ or $Z_2$, depending on whether they contribute to charged lepton Yukawa coupling or neutrino Yukawa coupling, respectively. 
\item 
shifts of the flavon VEVs. Thet may be resulted from the coupling between different flavons in the potential or interference by other field. The shift of the flavon VEV will not only contribute to lepton Yukawa couplings, but also modify the flavon masses and mixing directly. 
\end{itemize}
These sources may not be independent of each other, and contribute to flavour mixing at the same time. 

\begin{figure}[h!]
\begin{center}
\includegraphics[width=1\textwidth]{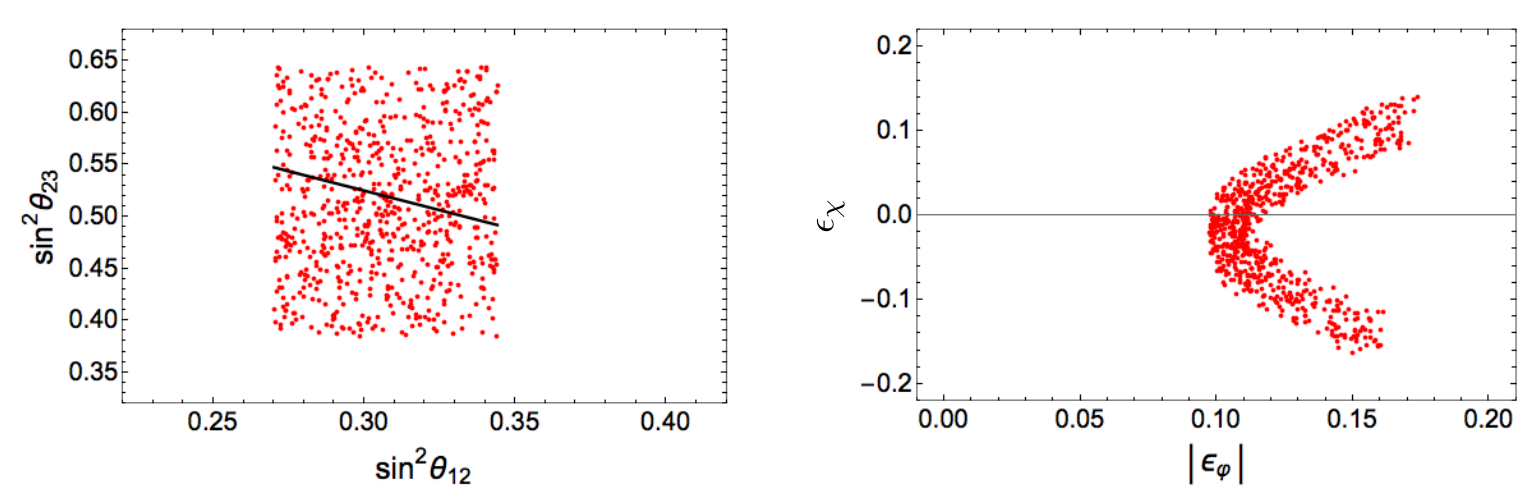}
\caption{Theoretical prediction of mixing angles (left panel) and the allowed parameter space of $|\epsilon_\varphi|$ vs $\epsilon_\chi$ (right panel). The straight line corresponds to $\epsilon_\chi=0$. }
\label{fig:mixing}
\end{center}
\vspace{-.5cm} 
\end{figure}

A very economical approach based on the second source is proposed in Ref. \cite{Pascoli:2016eld}. In this approach, flavon cross couplings between $\varphi$ and $\chi$ break the residual symmetries, shift the VEVs from Eq.~\eqref{eq:VEVs1} to 
\begin{eqnarray}
\langle \varphi \rangle \propto (1, \epsilon_{\varphi}, \epsilon_{\varphi}^*)^T \,,\quad
\langle \chi \rangle \propto (1-2 \epsilon_\chi, 1+\epsilon_\chi, 1+\epsilon_\chi)^T \,, 
\label{eq:flavon_vevs}
\end{eqnarray}
Here, $\epsilon_\varphi$ and $\epsilon_\chi$ are small parameters with $\epsilon_\varphi$ being complex and $\epsilon_\chi$ real. 
The PMNS matrix is approximately given by $U_\text{PMNS} = U_l^\dag(\epsilon_\varphi) U_\text{TBM} U_\nu(\epsilon_\chi)$ with $U_l(\epsilon_\varphi)$ and $U_\nu(\epsilon_\chi)$ representing $Z_3$- and $Z_2$-breaking effects, characterised by $\epsilon_\varphi$ and $\epsilon_\chi$, respectively. 
Deviations of mixing parameters from those in TBM are expressed as 
\begin{eqnarray} 
\sin\theta_{13}  \approx \sqrt{2}|\text{Im}(\epsilon_\varphi)|,~
\sin\theta_{12} \approx \frac{1-2\text{Re}(\epsilon_\varphi)+2\epsilon_\chi}{\sqrt{3}},~
\sin\theta_{23} \approx \frac{1+\text{Re}(\epsilon_\varphi)}{\sqrt{2}} \,,
\label{eq:mixing_angles}
\end{eqnarray} 
Furthermore, a sum rule between the mixing angle $\theta_{13}$ and CP-violating phase, 
\begin{eqnarray}
\delta \approx \mp( 90^\circ + \sqrt{2} \theta_{13})
\label{eq:sumrule_ra}
\end{eqnarray}  
for $\text{Im}(\epsilon_\varphi)>,<0$, respectively, is obtained. I show the numerical results for the correlation between $\theta_{12}$ and $\theta_{23}$ and the allowed parameter space of $|\epsilon_\varphi|$ vs $\epsilon_\chi$ in  Fig. \ref{fig:mixing}, where $3\sigma$ range data of mixing angles in \cite{globalfit} have been used. 

\section{CLFV induced by flavons}

The fact that neutrinos have masses and leptons mix is a convincing evidence of new physics. If there is a mechanism which can explain leptonic flavour mixing, it may also contribute to CLFV since neutrinos and left-handed charged leptons are unified in the electroweak symmetry. 

There are already some papers in the literature discussing the connections between CLFV and flavour symmetries. These papers have analysed the following contributions within flavour models. Higher dimensional operators which are not forbidden by $A_4$ \cite{Feruglio:2008ht}. Contributions of superpartners of leptons or flavons, since most of the flavour models are built in the framework of supersymmetry \cite{Feruglio:2008ht, Altarelli:2012bn, Muramatsu:2016bda}. Flavons as $SU(2)_L$-doublet Higgs \cite{Toorop:2010ex, Heeck:2014qea, Varzielas:2015joa, Kobayashi:2015gwa} and KK modes from warped flavour models \cite{Kadosh:2013nra} have also been discussed in some sence. 

I will pay more attention to the essential contribution of the flavour symmetry to CLFV. Here are my guiding principles: 
\begin{itemize}
\item
Simplicity. Only SM fields and gauge-invariant flavons will be included. Extra degrees of freedom not essential for explaining flavour mixing will be avoided. 
\item
Rigour. Non-trivial properties of the 3-dimensional Altarelli-Feruglio representation of $A_4$ will be taken care. 
\item
Consistency with data. To be consistent with oscillation data, NLO corrections to the mixing will be specified. How these corrections contribute to and connect with CLFV will be analysed carefully. 
\item
Testability. Unique features that can distinguish CLFV induced by non-Abelian discrete symmetry from other new physics will be emphasised.
\end{itemize}
These considerations lead us to the contribution only from the minimal extension of the SM that can explain oscillation data, namely, those flavons and necessary couplings with leptons. As $Z_3$ is a roughly preserved symmetry in the charged lepton sector, I will classify them into two parts: those consistent with the $Z_3$ residual symmetry and those contradicting it.

\subsection{$Z_3$-perserving channels}

From the Lagrangian terms in Eq.~\eqref{eq:Yukawa_coupling}, we can write out the couplings between flavons and charged leptons explicitly as 
\begin{eqnarray}
\mathcal{L}_l^\text{eff}&=&\frac{m_e}{v_\varphi} \left( \,\overline{e_L} e_R\, \varphi_1 + \overline{\mu_L} e_R \varphi_2 +\overline{\tau_L} e_R \varphi_2^* \right) \nonumber\\
&+&\frac{m_\mu}{v_\varphi} \left( \overline{\mu_L} \mu_R \varphi_1 + \overline{\tau_L} \mu_R \varphi_2 +\overline{e_L} \mu_R \varphi_2^* \right) \nonumber\\
&+&\frac{m_\tau}{v_\varphi} \left( \,\overline{\tau_L} \tau_R \, \varphi_1 + \, \overline{e_L} \tau_R \varphi_2 +\overline{\mu_L} \tau_R \varphi_2^* \right) + \text{h.c.}\,.
\label{eq:Feynman}
\end{eqnarray}
The $Z_3$ symmetry corresponds to the invariance under the transformation 
\begin{eqnarray} 
&&(e_{L,R}, \,\varphi_1) \,\to ~~~(e_{L,R}, \,\varphi_1) \,,\nonumber\\
&&(\mu_{L,R}, \,\varphi_2) \to \omega^2 (\mu_{L,R}, \,\varphi_2)\,,\nonumber\\
&&(\tau_{L,R}, \,\varphi_2^*) \,\to \omega~\, (\tau_{L,R},\, \varphi_2^*) \,,
\end{eqnarray} 
where $\omega=e^{i2\pi/3}$. 
Namely, $e_{L,R}$, $\varphi_1$ are invariant under the transformation of $Z_3$, $\mu_{L,R}$, $\varphi_2$ are covariant with a $Z_3$ charge 2, and $\tau_{L,R}$ are covariant with a $Z_3$ charge $1$. 
While the $Z_3$-invariant flavon $\varphi_1$ induces flavour-conserving processes, the $Z_3$-covariant flavon $\varphi_2$ is the main source of CLFV. 
As $e$, $\mu$ and $\tau$ take different $Z_3$ charges, it is easy to prove that the only allowed processes are $\tau^-\to \mu^+ e^-e^-$ and $\tau^-\to e^+ \mu^- \mu^-$. The other 3-body decay and all radiative decay modes are forbidden by the $Z_3$ symmetry \cite{triality}. 

In the case of transfer momentum much lower than the scale of flavour symmetry and flavon masses, one can integrate out $\varphi_{1}$ and $\varphi_2$, and derive the effective 4-fermion interactions. 
Those for $\tau^-\to \mu^+ e^-e^- $ and $\tau^- \to e^+ \mu^-\mu^-$ can be expressed as 
\begin{eqnarray}
\frac{m_\mu m_\tau}{v_\varphi^2m_{\varphi_2}^2} (\overline{e_L}\mu_R)(\overline{e_L}\tau_R) \,,\quad
\frac{m_\mu m_\tau}{v_\varphi^2m_{\varphi_2}^2} (\overline{\mu_R}e_L)(\overline{\mu_L}\tau_R) \,,
\label{eq:CLFV_Z3_ModelI}
\end{eqnarray}
respectively. The coefficients are the same, from which we get approximatively equal branching ratios of these two processes
\begin{eqnarray}
\text{Br}(\tau^-\to \mu^+ e^-e^-) \approx \text{Br}(\tau^-\to e^+ \mu^- \mu^-) \,, 
\label{eq:branching_sum_rule1}
\end{eqnarray}
both suppressed by $\big(\frac{m_\mu m_\tau v^2}{m_{\varphi_2}^2v_\varphi^2}\big)^2$. 
Assuming flavon VEV and flavon mass around the electroweak scale, the branching ratios are still two orders of magnitude below current experimental upper limit \cite{Pascoli:2016wlt}. 

\subsection{$Z_3$-breaking channels} 

Then we consider $Z_3$-breaking CLFV channels. Since the $Z_3$-breaking effect can give rise to non-zero $\theta_{13}$ and possible deviations of the other mixing parameters from their leading results,  
this effect should also be included in the discussion of CLFV. 
As there may be different $Z_3$-breaking origins, it is hard to do a generic analysis for the $Z_3$-breaking CLFV processes. 
To simplify the discussion, I will consider, as in \cite{Pascoli:2016wlt}, the $Z_3$-breaking from only flavon cross couplings. 

The breaking of $Z_3$ contributing to CLFV can be distinguished into three parts: 
\begin{itemize} 
\item the mixing of left-handed charged leptons $e_L$, $\mu_L$ and $\tau_L$, i.e., $U_l(\epsilon_\varphi)$\footnote{There is also a small mixing of right-handed charged leptons $e_R$, $\mu_R$ and $\tau_R$. it is suppressed by both $\epsilon_\varphi$ and the hierarchy of charged lepton masses due to arrangements of these particles as singlets in $A_4$, i.e., $e_R,\mu_R,\tau_R\sim \mathbf{1},\mathbf{1}',\mathbf{1}''$.}. 

\item the mixing between $Z_3$-invariant flavon $\varphi_1$ and $Z_3$-covariant one $\varphi_2$. 

\item the mass splitting between the two real degrees of freedom of $\varphi_2$.  
\end{itemize}
We put these comtributions into CLFV 3-body and radiative decay processes, and find all these decay modes are allowed. Compared with the $Z_3$-preserving processes, they are further suppressed by the additional parameter $\epsilon_\varphi$. 

For the 3-body decay processes, branching ratio sum rules of $\tau$ decays are observed: \begin{eqnarray}
& 2(B_{\mu^+\mu^-e^-}-2B_{\mu^+\mu^-\mu^-})^2+(5B_{e^+e^-\mu^-}+10B_{\mu^+\mu^-\mu^-}-6B_{\mu^+\mu^-e^-})B_{e^+e^-\mu^-}=0 \,,\nonumber\\
& B_{e^+e^-\mu^-} \approx 8|\epsilon_\varphi|^2 \text{Br}(\tau^-\to \mu^+ e^- e^-) \,.
\label{eq:branching_sum_rule2}
\end{eqnarray}
Here, $B_{\mu^+\mu^-e^-}$, $B_{\mu^+\mu^-\mu^-}$, $B_{e^+e^-\mu^-}$ are branching ratios of $\tau^-\to \mu^+\mu^-e^-$, $\tau^-\to \mu^+\mu^-\mu^-$ and $\tau^-\to e^+ e^- \mu^- $, respectively. In the limit $m_{\varphi_1}\ll m_{\varphi_2}$, we get $B_{\mu^+\mu^-e^-}\approx 2 B_{\mu^+\mu^-\mu^-} \gg B_{e^+e^-\mu^-}$, and on the contrary, we obtain $B_{\mu^+\mu^-e^-} \approx 4 B_{\mu^+\mu^-\mu^-} \approx 2 B_{e^+e^-\mu^-}$.  These processes are weaker than the $Z_3$-preserving processes because of the $\epsilon_\varphi$ suppression. For $\tau^- \to e^+e^-e^-$ and $\mu^- \to e^+e^-e^-$, their branching ratios are suppressed by electron mass, thus, far away from experimental limit.

Radiative decays are also allowed by cross couplings. $\tau^- \to \mu^- \gamma$ and $\tau^- \to e^- \gamma$ are induced by the mixing between charged lepton flavour eigenstates and by the mixing between $\varphi_1$ and $\varphi_2$. The approximately equal branching ratios are predicted,
\begin{eqnarray}
\text{Br}(\tau^- \to e^- \gamma) \approx \text{Br}(\tau^-\to \mu^- \gamma) \,.
\label{eq:branching_sum_rule3}
\end{eqnarray}
Assuming the flavon VEV and masses above the electroweak scale, these branching ratios $\lesssim10^{-11}$, 3 orders of magnitude below the current best experimental limits. 

\begin{figure}[h!]
\begin{center}
\includegraphics[width=.7\textwidth]{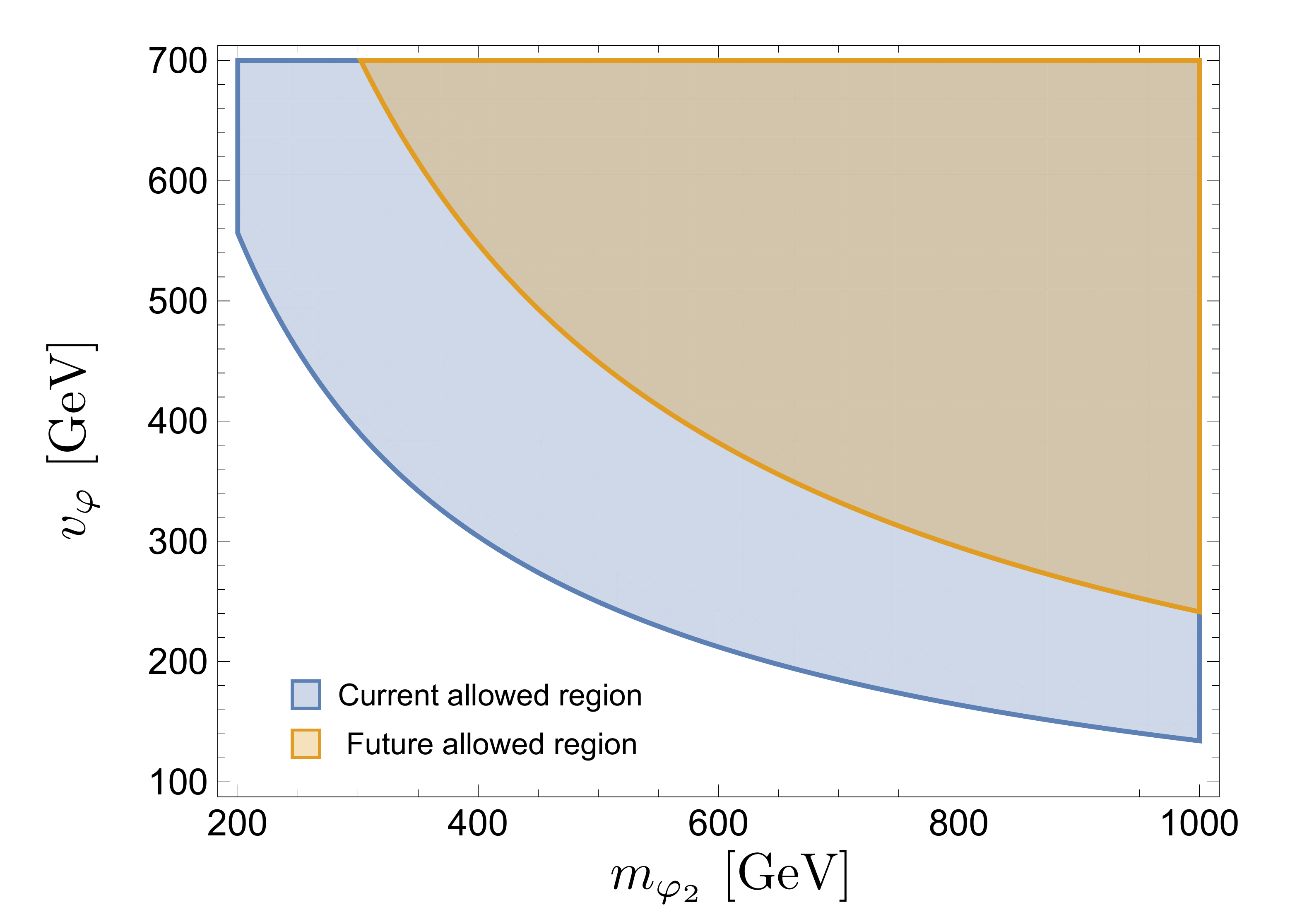}
\caption{Flavon VEV and $\varphi_2$ mass constrained by $\mu^-\to e^- \gamma$ experiments. $|\epsilon_\varphi|$ is fixed at 0.1 for generating $\theta_{13}$. The current (MEG) and future (MEG II) constraints are set to be $\text{Br}(\mu^-\to e^-\gamma)<4.2\times 10^{-13}$ and $4\times 10^{-14}$, respectively. }
\label{fig:CLFV}
\end{center}
\vspace{-.5cm} 
\end{figure}

The most stringent constraint is from $\mu^- \to e^- \gamma$. It is induced by the mixing of charged leptons and the mass splitting of the two real degrees of freedom of the complex flavon $\varphi_2$. This process is suppressed by $\epsilon_\varphi$ and $m_\mu$, but it has been measured more precisely than the $\tau$ decay. Regions of $v_\varphi$ and $m_{\varphi_2}$ allowed by current experiments and testable at the near future experiments are shown in Fig. \ref{fig:CLFV}. The current upper limit is given by the MEG experiment, around $4.2 \times 10^{-13}$ (90\% CL) \cite{MEG}. By fixing the flavon VEV at the electroweak scale, we obtain the lower limit of the $\varphi_2$ mass is around 500 GeV.  In the future, MEG II will push the upper limit of the branching ratio to $4\times 10^{-14}$ \cite{MEG2}. This experiment will have the potential to exclude a large parameter space.

\section{Conclusion}

Flavour symmetries are usually treated as the origin of leptonic flavour mixing. The newly introduced interactions in flavour models will not only generate flavour structure in lepton mass matrices, but also contribute to CLFV processes. The flavon-triggered CLFV has strong connections with the symmetries. 

In $A_4$ models,  all CLFV processes can be classified by the residual symmetry $Z_3$. The only allowed $Z_3$-preserving processes are $\tau^- \to \mu^+ e^- e^-$, $\tau^- \to e^+ \mu^- \mu^-$ and their conjugate processes. All the other 3-body and radiative decays are $Z_3$-breaking. Several CLFV sum rules are obtained. This is a highlighted feature to connect non-Abelian discrete flavour symmetries with CLFV. While the $Z_3$-preserving processes are suppressed by charged lepton masses, the $Z_3$-breaking processes are further suppressed due to the consistency with oscillation data. The current experimental constraints are loose. Hundreds of GeV scale flavon VEV and mass are still allowed.

\Acknowledgements
I am grateful to the NuPhys 2016 organizers for their kind invitation.

\end{document}